\documentclass[12pt,a4paper]{iopart}

\usepackage{amsfonts,iopams,cite}
\usepackage{mathrsfs}
\usepackage{graphicx}
\usepackage{dcolumn}  
\usepackage{psfrag}

\newcommand{\id}{\hat{1}}
\newcommand{\im}{i}

\newcommand{\meanv}[1]{\big\langle #1 \big\rangle}
\newcommand{\var}[1]{\left( \Delta #1\right)^2}
\newcommand{\varsb}[1]{\left[ \Delta #1\right]^2}

\newcommand{\ot}{\otimes}
\newcommand{\proj}[1]{\vert #1 \rangle\!\langle #1 \vert}

\newcommand{\ch}{\mathrm{ch}}
\newcommand{\sh}{\mathrm{sh}}

\begin{document}

\title{Beyond pure state entanglement for atomic ensembles}
\date{\today}
\author{Julia Stasi\'nska$^{1,\#}$, Simone Paganelli$^1$
and Anna Sanpera$^{2,1}$}
\address{$^1$ Departament de F\'isica. Grup de F\'isica Te\`orica: Informaci\'o
i Fen\`omens Qu\`antics,\\
Universitat Aut\`onoma de Barcelona, 08193 Bellaterra (Barcelona), Spain.}
\address{$^2$ ICREA-Instituci\'o Catalana de Recerca i Estudis Avan\c cats,
Spain.}
\eads{ \mailto{$^{\#}$ julsta@ifae.es}}

\begin{abstract}
We analyze multipartite entanglement between atomic ensembles within
quantum matter-light interfaces. In our proposal, a polarized light beam crosses
sequentially several polarized atomic ensembles impinging on each of them at a
given angle $\alpha_{i}$. These angles are crucial parameters  for shaping the
entanglement since they are directly connected to the appropriate combinations
of the collective atomic spins that are squeezed.  We exploit such scheme to go
beyond the pure state paradigm proposing realistic experimental settings to
address multipartite mixed state entanglement in continuous
variables.
\end{abstract}



\section{Introduction}\label{sec:introduction}

Experiments exploiting the quantum character of atom-light interactions are
often carried out in the regime of cavity QED where, due to cavity, the
coupling between individual atoms and photons, represented by the optical depth
is highly enhanced. An alternative approach to achieve an efficient atom-photon
coupling are atomic ensembles, where large optical depths are obtained due to
to the macroscopic or mesoscopic number of atoms of the sample, as proposed
initially by Kuzmich and coworkers \cite{Kuzmich1997_spinsq}.

Atomic ensembles typically refer to samples composed of $10^6$ to $10^{12}$
atoms. In the atomic ensembles the spin of each atom, $\hat{\bf{j}}_{i}$,
originating either from its hyperfine structure or from its nuclear spin (e.g.
alkaline-earth fermions) can be disregarded and the sample can be characterized
by its collective
spin
$\hat{\bf{J}}=(\hat{J}_{x},\hat{J}_{y},\hat{J}_{z})=\sum_{i}\hat{\bf{j}}_{i}$.
If the sample is polarized, the component of $\hat{\bf{J}}$ along the
polarization axis can be approximated to a c-number, let's say
$\hat{J}_{x}=\sum_{i}\hat j_{x}^{i}\simeq \langle \hat{J}_{x}\rangle$, while the
orthogonal components behave as conjugate canonical
variables$[\hat{J}_{y},\hat{J}_{z}]=i\hbar \langle\hat{J}_{x}\rangle$. Such an
elegant and simple description of polarized atomic ensembles makes them
extremely appealing for quantum information processing in the domain of
continuous variables.

The interaction between a polarized atomic ensemble and an off-resonant linearly
polarized laser beam results -- at the classical level -- in a Faraday
rotation. The latter refers to the rotation experienced by the polarization of
light when it propagates through a magnetic medium, in our case the polarized
atomic ensemble. Furthermore, this interaction leads to a quantum interface,
i.e., an exchange of quantum fluctuations between the matter and the light.
Seminal results exploiting such interface are the generation of atomic spin
squeezed states produced by the interaction of coherent atomic sample and a
squeezed light beam~\cite{Kuzmich1997_spinsq,Kuzmich2000_spinsq_experiment}, the
realization of the atomic quantum memory~\cite{Julsgaard2004_memory} and the
establishing of entangled state of two spatially
separated atomic ensembles~\cite{Julsgaard2001_entanglement}. The above
situations require a final projective measurement on the light (homodyne
detection) that projects the atomic ensembles into their desired final state.
This "measurement induced" mapping of fluctuations between different physical
systems provides a powerful tool to design quantum correlations. The potential
applications of such a genuine quantum tool have just started. Among them are
the application in
quantum metrology~\cite{Napolitano2010NJP_metrology,
Napolitano2011Nature_metrology},
quantum magnetometry~\cite{Wolfgramm2010_magnetometry} and quantum spectroscopy
for detection of magnetic ordering in spinor Bose-Einstein
condensates~\cite{Eckert2007_spinor}, or strongly correlated spin systems
realized with ultracold atomic
gases~\cite{Eckert2007_nphys,Roscilde2009_polspectr,deChiara2010,
Esselinger2011_PRL}.

A milestone achievement using quantum matter-light interfaces was the
generation of entanglement between two spatially separated atomic
ensembles~\cite{Julsgaard2001_entanglement} mediated by a single light beam
propagating sequentially through both of them. In such setup, the entanglement
between the atomic ensembles was established as soon as the light was measured,
independently of the outcome of the measurement. However the detection of the
entanglement thus generated, based on variance
inequalities criteria~\cite{Duan2000_sepCV,vanLoock2003_sepCVmult}, is rather
challenging. As shown in~\cite{Julsgaard2001_entanglement} its verification can
be made feasible with the help of external magnetic fields applied to the
ensembles. Nonetheless, it has been shown in \cite{Stasinska2009_meso} that
entanglement verification without external magnetic fields is also possible if
the light crosses through each sample at a given angle $\alpha_i$. In this
setup, that we name "geometrical", the incident angles play a critical role
since they can be used to shape both quantitatively and qualitatively, the
entanglement between the atomic ensembles.

The geometrical scheme of entangling different atomic systems using quantum
interfaces can be extended beyond the bipartite pure state setting to include
both several parties and noise. Multipartite mixed states entanglement, has been
mainly  addressed for finite dimensional Hilbert spaces and it is
poorly understood in all but the simplest situations. For instance, recent
results suggesting the potential supremacy of noisy quantum computation versus
pure state multipartite quantum
computation\cite{Bruss2011_multipart,Memarzadeh2011, Bremner2009_useful} are
highly
intriguing. For continuous variables even less is known. Atomic-ensemble systems
belong to the important class of continuous-variable states known as Gaussian,
whose characterization and mathematical description is much simpler.
For instance, the complete classification of three-mode Gaussian systems has
been provided in \cite{Dur1999_sepmulti,Giedke2001_Gauss3mode}. However, even in
the simplest multipartite Gaussian scenario, the verification of the
entanglement using variance inequalities~\cite{vanLoock2003_sepCVmult}
becomes a very intricate task in the atomic-ensemble setup. Here we address the
generation and detection of multipartite entanglement in Gaussian mixed states
in experimentally feasible systems using the geometrical scheme.  In particular,
we provide physical bounds on atomic ensembles' parameters that lead to
generation of bound entanglement, providing a guideline on their experimental
accessibility. Bound entanglement refers to the entanglement which cannot be
distilled, i.e., no pure state entanglement can be obtained from it by means of
local operations and classical communication (LOCC). In the multipartite
setting, a state can be bound entangled with respect to certain parties while
distillable with respect to others. Although bound entanglement may seem useless
for quantum information processing, it has been shown that for discrete systems
it can be used for certain tasks (e.g. channel discrimination) or activated
\cite{Vollbrecht2002,Horodecki2005,Augusiak2006_Smolin,Masanes2006,Masanes2008,
Piani2009}.

A particular example of bound entanglement  we are going to deal with is the
so-called Smolin state \cite{Smolin2001_be}. In the discrete case, this state
corresponds to a four qubit mixed state formed by an equal-weight mixture of
Bell states that has the property that its entanglement can be unlocked.
Moreover, it has been shown that it violates maximally a Bell inequality, being
at the same time useless for secrete key distribution
\cite{Augusiak2006_Smolin}. The Smolin state has been very recently addressed
mathematically within the continuous variables stabilizer
formalism\cite{Zhang2011_Smolin}. Here we will perform its covariance matrix
analysis that gives more insight into the nature of the state and demonstrate
how such type of states can be experimentally realized using quantum light
interfaces with atomic-ensembles.

Since both the atomic ensembles and the light assume a unified
continuous-variable description and given the inherent Gaussian character of the
setting we are considering, we use a covariance matrix
formalism~\cite{Sherson2005_Gauss1} to describe in a compact way the generation
and manipulation of entanglement. Such a formalism, as we shall see, enlightens
the possibilities offered by atomic samples as a quantum toolbox to generate
many different types of continuous variables entangled states in an
experimentally accessible way.

The paper is organized as follows. In Section \ref{sec:CV_sys} we review, for
completeness, the description of matter and light as continuous
variables systems, and describe the matter-light interaction in terms of an
effective Hamiltonian and the corresponding evolution equations. Then we derive
explicitly all the steps of the matter-light interface in terms of covariance
matrices, symplectic transformations and Gaussian operations. In Section
\ref{sec:mixed_entanglement} we analyze the properties of the interface leading
to mixed state entanglement with atomic ensembles. First, we consider the
bipartite case in Section \ref{subs:bipartite}, setting the bounds on system
parameters that produce mixed state entanglement. In Section
\ref{subs:multipartite} we move to multipartite {\it bound} entanglement in
continuous variables. Here, we consider a general frame in which initially the
atomic samples are in a thermal state with different amounts of noise. This
allows us to set physical bounds on the generation of such type of entanglement
in the CV scenario. In Section \ref{subs:smolin} we analyze the much more
involved Smolin state\cite{Zhang2011_Smolin}. We show that the geometrical setup
we propose provides all the tools needed to generate and unlock the Smolin
state, including a (quantum) random number generator. Finally, in Section
\ref{sec:summary} we present our conclusions.

\section{Continuous Variables systems}\label{sec:CV_sys}
\subsection{The Faraday Interaction}\label{subs:Faraday_int}

The setting we are considering consists of several atomic ensembles and
light beams, the latter playing the role of information carriers between the
atomic samples. At a time, only a single light beam interacts with the atomic
ensembles. In previous proposals, the light beam after interacting with the
atomic ensembles was measured, inducing entanglement between the different
samples (see
e.g.\cite{Sherson2005_Gauss1,Madsen2004_Gauss,Julsgaard2001_entanglement,
Stasinska2009_meso}). Here, we will have a closer look at the situation in which
the light beam is not measured after the interaction. Such action, as we will
see, acts as a truly quantum Gaussian random number generator. We will show
that through such procedure one can generate multipartite mixed states, in
particular bound entangled states.

As pointed out in the introduction, each atomic ensemble is described by its
collective angular momentum
$\mathbf{\hat{J}}=(\hat{J}_{x},\hat{J}_{y},\hat{J}_{z})$. Atoms are assumed to
be all polarized along the $x$ direction (e.g. prepared in a particular
hyperfine state) so that fluctuations in the $\hat{J}_x$ component of the
collective spin are very low and this variable can be treated as a classical
number $\hat{J}_x\approx \meanv{\hat{J}_x}\equiv \hbar J_x=\hbar N_{\mathrm{at}}
j$. By appropriate normalization the orthogonal spin components are made to
fulfill the canonical commutation relation, $\left[\hat{J}_y/\sqrt{\hbar
J_x},\hat{J}_z/\sqrt{\hbar J_x}\right]=i\hbar $. Notice that they have non-zero
fluctuations. To stress the continuous variable character of the system, we rename
the above variables as ``position'' and ``momentum'' :
\begin{equation}\label{canonical_at}
\hat{x}_{A}=\frac{\hat{J}_{y}}{\sqrt{\hbar J_{x}}},\mbox{\hspace{2cm}}
\hat{p}_{A}=\frac{\hat{J}_{z}}{\sqrt{\hbar J_{x}}}.
\end{equation}

From now on we will use only the canonical variables $\hat{x}_{A}, \hat{p}_{A}$
to refer to the atomic sample, where the subindex $A$ stands for atomic
ensemble. Later on when dealing with different atomic ensembles the notation
$\hat{x}_{A,n}, \hat{p}_{A,n}$, when we refer to the $n$th atomic sample, will
be used.

Light is taken to be out of resonance from any relevant atomic transition and
linearly polarized along the $x$-direction. We use the Stokes description
$\mathbf{\hat{s}}=(\hat{s}_x,\hat{s}_y,\hat{s}_z)$ for the light polarization.
The components $\hat{s}_k$ $(k=x,y,z)$ correspond to the differences between the
number of photons (per unit time) with $x$ and $y$ linear polarizations, $\pm
\pi/4$ linear polarizations and the two circular polarizations, i.e.,

\begin{equation}\label{stokes}
\begin{array}{l}
\displaystyle \hat{s}_x=\frac{\hbar}{2} (\hat{n}_x-\hat{n}_y)=\frac{\hbar}{2}
(\hat a_x^\dag\hat a_x-\hat a_y^\dag\hat a_y), \\[2ex]
\displaystyle \hat{s}_y=\frac{\hbar}{2}
(\hat{n}_{\nearrow}-\hat{n}_{\searrow})=\frac{\hbar}{2}(\hat a_x^\dag\hat
a_y+\hat a_y^\dag\hat a_x),\\[2ex]
\displaystyle
\hat{s}_z=\frac{\hbar}{2}(\hat{n}_{\circlearrowleft}-\hat{n}_{\circlearrowright}
)=\frac{\hbar}{2
\im}(\hat a_x^\dag\hat a_y-\hat a_y^\dag\hat a_x).
\end{array}
\end{equation}

The Stockes operators are well suited for the microscopic description of
interaction with atoms, however, effectively only the following macroscopic
observables will be relevant: $\hat{S}_k=\int_{0}^{T} \hat{s}_k(t)\mathrm{d}t$,
where $T$ is the duration of the light pulse. The so defined operators obey
standard angular momentum commutation rules. The assumption of linear
polarization along direction $x$ allows for the approximation
$\hat{S}_x\approx\meanv{\hat{S}_x}\equiv N_{ph} \hbar/2 $. Once more, the
remaining orthogonal components $\hat{S}_y$ and $\hat{S}_z$ are appropriately
rescaled in order to make them fulfill the canonical commutation rule,
$\left[\hat{S}_y/\sqrt{\hbar S_x},\hat{S}_z/\sqrt{\hbar S_x}\right]=\im \hbar$.
Straightforwardly, an equivalent equation to equation (\ref{canonical_at})
arises:
\begin{equation}\label{canonical_light}
\hat{x}_{L}=\frac{\hat{S}_{y}}{\sqrt{\hbar S_{x}}},\mbox{\hspace{2cm}}
\hat{p}_{L}=\frac{\hat{S}_{z}}{\sqrt{\hbar S_{x}}},
\end{equation}
which allows to treat the light polarization degrees of freedom on the same
footing as the atomic variables.

In the situation in which a light beam propagates in the $YZ$
plain and passes through a single ensemble at angle $\alpha$ with
respect to direction $z$, the atom-light interaction can be
approximated to the following QND effective Hamiltonian (see
\cite{Hammerer2010_rev} and references therein for a detailed
derivation):

\begin{equation}\label{hamiltonian}
\hat{H}_{\mathrm{int}}^{\mathrm{eff}}(\alpha)=-\kappa\hat{p}_{L}
(\hat{p}_{A} \cos{\alpha}+\hat{x}_{A} \sin{\alpha}).
\end{equation}
The parameter $\kappa$ is the coupling constant with the dimension of the
inverse of an action. Notice that such Hamiltonian leads to a bilinear coupling
between the Stokes operator and the collective atomic spin operators. Evolution
can be calculated through the Heisenberg equation for the atoms and using
Maxwell-Bloch equation for light, neglecting retardation effects. The variables
characterizing both systems (atom and light) transform according to the
following equations (\cite{Hammerer2010_rev} and references therein):
\numparts
\begin{eqnarray}
\hat{x}_{A}^{\mathrm{out}}&=&\hat{x}_{A}^{\mathrm{in}}-\kappa
\hat{p}_L^{\mathrm{in}} \cos \alpha,\label{propagationa}\\
\hat{p}_{A}^{\mathrm{out}}&=&\hat{p}_{A}^{\mathrm{in}}+\kappa
\hat{p}_L^{\mathrm{in}} \sin
\alpha,\label{propagationb}\\
\hat{x}_L^{\mathrm{out}}&=&\hat{x}_L^{\mathrm{in}}
-\kappa (\hat{p}_{A}^{\mathrm{in}} \cos{\alpha}+\hat{x}_{A}^{\mathrm{in}}
\sin{\alpha}),\label{propagationc}\\
\hat{p}_L^{\mathrm{out}}&=&\hat{p}_L^{\mathrm{in}}.\label{propagationd}
\end{eqnarray}
\endnumparts
where the subscript in/out refers to the variable before/after the interaction.
The above equations can be generalized to the case in which a single light beam
$(\hat{x}_L,\hat{p}_L)$ propagates through many samples shining at the $n$th
sample at a certain angle $\alpha_n$. A complete description taking into account
inhomogeneities of the atomic samples or the light beams has been considered in
\cite{Koschorreck2009}.

Due to the strong polarization constraint, both the atomic ensembles and the
light are initially Gaussian modes. Moreover, the Hamiltonian is linear in both
atomic and light quadratures, and therefore quadratic in creation and
annihilation operators. Such interaction Hamiltonians correspond to Gaussian
transformations preserving the Gaussianity of the input state. These facts
enable us to tackle the quantum atom-light interface within a covariance matrix
formalism.

\subsection{ The atom light interface in the covariance matrix
formalism}\label{subs:CM_formalism}

We start by reviewing the most basic concepts needed to describe Gaussian
continuous-variable systems. For further reading, the reader is referred to
\cite{Giedke2002_gaussops,Braunstein2005_rev,Adesso2007_rev} and references
therein. For a general quantum system of $N$ pairs of canonical degrees of
freedom (``position'' and ``momentum''), the commutation relations fulfilled by
the canonical coordinates $\hat{R}=(\hat{x}_1, \hat{p}_1, \ldots, \hat{x}_{N},
\hat{p}_{N})$ can be represented in a matrix form by the symplectic matrix
$\mathcal{J}_N: [\hat{R}_i,\hat{R}_j]=\im \hbar (\mathcal{J}_N)_{ij}$,
$i,j=1,\ldots, 2N$, where
\begin{equation}\label{sympl_J}
\mathcal{J}_{N}=\bigoplus_{\mu=1}^{N} \mathcal{J},\quad \mathcal{J}=\left(
                                                                      \begin{array}{cc}
                                                                        0 & 1 \\
                                                                        -1 & 0 \\
                                                                      \end{array}
                                                                    \right).
\end{equation}
Gaussian states are, by definition, fully described by the first and second
moments of the canonical coordinates. Hence, rather than describing them by
their infinite-dimensional density matrix $\varrho$, one can use the Wigner
function representation
\begin{equation}\label{wigner_gauss}
W(\zeta)=\frac{1}{\pi^N \sqrt{\det \gamma}} \exp \left[-(\zeta-d)^T \gamma^{-1} (\zeta-d)\right],
\end{equation}
which is a function of the first moments through the displacement vector $d$,
and of the second moments through the covariance matrix $\gamma$, defined as:
\begin{equation}\label{disp_cov_def}
d_i=\Tr (\varrho \hat{R}_i),\qquad \gamma_{ij}=\Tr (\varrho
\{\hat{R}_i-d_i,\hat{R}_j-d_j\}).
\end{equation}
The variable $\zeta=(x_1,p_1,\ldots,x_N,p_N)$ is a real phase space vector with
probability distribution given by the Wigner function. The covariance matrix
corresponding to a quantum state must fulfill the positivity condition
\begin{equation}\label{positivity}
\gamma+\im \mathcal{J}_N\geq 0.
\end{equation}
In the particular case of a physical system consisting of several atomic
ensembles and single light beam the most general covariance matrix takes the
form
\begin{equation}\label{cm_ABC}
\gamma=\left(
\begin{array}{cc}
  \gamma^{A} & C \\
  C^T & \gamma^{L}
\end{array}\right),
\end{equation}
where the submatrix $\gamma^L$ corresponds to the light mode, $\gamma^A$ to
the atomic ensembles, that initially
reads $\gamma^A_{in}=\gamma^{A_{1}}_{in}\oplus\cdots\oplus\gamma^{A_{n}}_{in}$
and $C$ accounts for the correlations between the atomic ensembles and the
light.

If a Gaussian state undergoes a unitary evolution preserving its Gaussian
character, as it is the case here, then the corresponding transformation at the
level of the covariance matrix is represented by a symplectic matrix $S$ acting
as
\begin{equation}\label{symplectic2}
\gamma_{\mathrm{out}}=S^{T}\gamma_{\mathrm{in}} S.
\end{equation}
Let us illustrate how to reconstruct the evolution of the covariance matrix
from the propagation equations (\ref{propagationa})--(\ref{propagationd}).
Notice that the variables describing the system after interaction (out) are
expressed as a linear combination of the initial ones (in). Let us denote this
linear transformation by $K$
\begin{equation}
K: (\hat x_{A,n}^{\rm{out}},\hat p_{A,n}^{\rm{out}},\hat
x_{L}^{\rm{out}},\hat p_{L}^{\rm{out}})^{T}=K (\hat x_{A,n}^{\rm{in}},\hat
p_{A,n}^{\rm{in}},\hat x_{L}^{\rm{in}},\hat p_{L}^{\rm{in}})^{T}.
\end{equation}
In our case, $K$ can be straightforwardly obtained from the evolution
equations (\ref{propagationa})--(\ref{propagationd}). For a single atomic mode
the transformation reads:
\begin{equation}
\left(
  \begin{array}{c}
    \hat{x}_A^{\mathrm{out}} \\
    \hat{p}_A^{\mathrm{out}} \\
    \hat{x}_L^{\mathrm{out}} \\
    \hat{p}_L^{\mathrm{out}}
  \end{array}
\right)= \left(
  \begin{array}{cccc}
    1 & 0& 0 & -\kappa \cos \alpha \\
    0 & 1 & 0 & \kappa \sin \alpha \\
    -\kappa \sin \alpha & -\kappa \cos \alpha & 1 & 0 \\
    0 & 0 & 0 & 1
  \end{array}
\right)\left(
  \begin{array}{c}
    \hat{x}_A^{\mathrm{in}} \\
    \hat{p}_A^{\mathrm{in}} \\
    \hat{x}_L^{\mathrm{in}} \\
    \hat{p}_L^{\mathrm{in}}
  \end{array}
\right).
\end{equation}

\noindent Since the interaction Hamiltonian is bilinear, the matrix $K$ can be
directly applied to a phase space vector $\zeta$ and correspondingly to the
covariance matrix, however the sign of the coupling constant $\kappa$ should be
changed. This is so because the phase space variables evolve according to the
Schr\"odinger picture, whereas the quadratures, being operators transform
according to the Heisenberg picture. Therefore, we
define $\tilde{K}=K\big|_{\kappa\to(-\kappa)}$, which we apply to the phase
space vector and covariance matrix as
\begin{equation}
\label{symplectic}
\zeta_{\mathrm{out}}^{T}\gamma_{\mathrm{in}}^{-1} \zeta_{\mathrm{out}}=\zeta_{\mathrm{in}}^T \tilde{K}^T \gamma_{\mathrm{in}}^{-1} \tilde{K} \zeta_{\mathrm{in}}
=\zeta_{\mathrm{in}}^T (\tilde{K}^{-1} \gamma_{\mathrm{in}} (\tilde{K}^{T})^{-1})^{-1} \zeta_{\mathrm{in}}
=\zeta_{\mathrm{in}}^{T} \gamma_{\mathrm{out}}^{-1}\zeta_{\mathrm{in}},
\end{equation}
leading to $S=(\tilde{K}^{T})^{-1}$. The above formalism has been
explicitly developed for a single sample and a single beam, but it easily
generalizes to an arbitrary number of atomic ensembles and  light beams, as well
as to different geometrical settings.

Finally, the last ingredient essential to describe the matter--light interface
at the level of the covariance matrix is the effect of the homodyne detection of
light \cite{Giedke2002_gaussops}. A homodyne measurement on the light
quadratures acts as a Gaussian map on the atomic covariance matrix. Assuming a
zero initial displacement and covariance matrix of the form (\ref{cm_ABC}), the
measurement of the quadrature $\hat{x}_L$ with outcome $\tilde{x}_L$ leaves the
atomic system in a state described by a covariance matrix
\cite{Eisert2002_Gaussimpossible,Giedke2002_gaussops, Sherson2005_Gauss1}.
\begin{equation}\label{cm_measurement1}
\gamma^{A'}=\gamma^{A}-C (X\gamma^{L}X)^{-1} C^T,
\end{equation}
and displacement
\begin{equation}\label{cm_measurement2}
d_{A}=C (X\gamma^{L}X)^{-1} (\tilde{x}_L,0),
\end{equation}
where the inverse is understood as an inverse on the support whenever the matrix
is not of full rank and $X$ is a two-dimensional diagonal matrix with diagonal
entries $(1,0)$. The measurement of the quadrature $\hat{p}_L$ affects the
system analogously. Let us note here that if the light is not measured after the
interaction, the state of the atomic sample is characterized by the covariance
matrix $\gamma^{A}$  obtained after tracing out the light degrees of freedom,
with the displacement being a Gaussian random variable according to equations
(\ref{propagationa}) and (\ref{propagationb})
\begin{equation}
d_A={-\kappa \bar{p}_L \cos \alpha, \kappa \bar{p}_L \sin \alpha},
\end{equation}
where $\bar p_L$ denotes a Gaussian random variable associated with momentum
operator.

An important step in the matter--light interface when several atomic ensembles
are present is the analysis of the quantum correlations between the different
atomic samples once the light beam has been measured. The symplectic formalism
provides the whole information about the atomic covariance matrix
and displacement vector after the interaction. This makes verification of
entanglement amenable to covariance matrix entanglement criteria (see also
\cite{Madsen2004_Gauss,Sherson2005_Gauss1}).

An operational separability criterion, i.e., state-independent, which can be
only applied when the full covariance matrix is available, is the positive
partial transposition (PPT) criterion \cite{Horodecki1996_crit, Peres1996_crit}.
For continuous variable systems, it corresponds to partial time reversal of
the covariance matrix \cite{Simon2000_pptCV}, i.e. a change of the sign of the
momentum for the chosen modes. If the partially time reversed covariance matrix
does not fulfill the positivity condition (\ref{positivity}), the corresponding
state is entangled. This test, however, checks only bipartite entanglement. For
Gaussian states this criterion is necessary and sufficient to detect
entanglement in all partitions of  $1\times N$ modes. In the multipartite
scenario, a state may be PPT with respect to all its bipartite divisions and
still not be fully separable. Such states are bound entangled states. For
three-mode Gaussian states, a operational separability criterion distinguishing
a fully separable state from a fully PPT entangled state was given in
\cite{Giedke2001_Gauss3mode}. We will use such criterion together with the PPT
criterion in section \ref{subs:multipartite} to demonstrate how different types
of entanglement arise in tripartite cluster-like states at finite temperature.

Experimentally, however, it is more convenient to check separability via
variances of the collective observables, originally proposed for two mode states
in \cite{Duan2000_sepCV} and generalized to many-mode states
in \cite{vanLoock2003_sepCVmult}. Such criteria states that if an $N$ mode state
is separable, then the sum of the variances of the following operators:
\begin{eqnarray}
\hat{u}=h_1 \hat{x}_{1}+\ldots+h_{N} \hat{x}_{N}\nonumber\\
\hat{v}=g_1 \hat{p}_{1}+\ldots+g_{N} \hat{p}_{N}
\end{eqnarray}
is bounded from below by a function of the
coefficients $h_1,\ldots,h_{N},g_1,\ldots,g_{N}$. Mathematically, the inequality
is expressed as
\begin{equation}\label{var_ineq}
\var{\hat{u}}+\var{\hat{v}}\geq f(h_1,\ldots,h_{N},g_1,\ldots,g_{N})\hbar,
\end{equation}
where
\begin{equation}
f(h_1,\ldots,h_{N},g_1,\ldots,g_{N})=\left|h_l g_l+\sum_{r\in
I}h_r g_r\right|+\left|h_m g_m+\sum_{s\in I'}h_s g_s\right|.
\end{equation}
In the above formula the two modes, $l$ and $m$, are distinguished and
the remaining ones are grouped into two disjoint sets $I$ and $I'$. The
criterion (\ref{var_ineq}) holds for all bipartite splittings of a state defined
by the sets of indices $\{l\}\cup I$ and $\{m\}\cup I'$. For two mode states,
the criterion becomes a necessary and sufficient entanglement test, however only
after the state is transformed into its standard form by local
operations \cite{Duan2000_sepCV}. This local transformations, however, are
determined by the form of the covariance matrix. In this sense, the knowledge of
the full covariance matrix is essential in order to determine whether the state
is entangled. Since in experiments usually one does not have access to the full
covariance matrix, one cannot assume that this criterion decides unambiguously
about separability.

\section{\bf Beyond the pure state entanglement}\label{sec:mixed_entanglement}

Here we analyze mixed state entanglement for atomic ensembles. We start
our analysis with bipartite states and show that the entanglement induced by the
measurement of light, despite its irreversible nature, can be erased by
making the samples interact with a second light beam, in a similar fashion as it
happens for the pure states \cite{Stasinska2009_meso}. We then address
multipartite entanglement and show how bound entanglement can be created in a
tripartite setting (cluster state) using thermal states. Finally, we analyze the
effect of randomness introduced in the multipartite setting by the action of the
light which interacts with all the atomic samples but is not measured. We show
that through such procedure one may produce unlockable bound entanglement.

\subsection{\bf Bipartite entanglement of thermal states.}\label{subs:bipartite}

Let us start with the setup in which the atomic samples are not in the
minimum-fluctuation coherent state (vacuum), but in a general thermal state.
Under such assumptions, the initial state of the composite system is given by
the following covariance matrix for atoms and light
\begin{equation}
\gamma_{\rm{in}}=n_1\id_{2}^{A}\oplus \cdots \oplus n_N\id_2^{A}\oplus\id_2^{L},
\end{equation}
where the identity  $\id_{2}$ stands for a single mode and parameters
$n_1,\ldots, n_N$ are related to temperature through $n_i=1/ \tanh[\beta_i
\omega /2 ]$ ($i=1,\cdots,N$), where $\beta$ is the inverse of the temperature,
and $\omega$ is the effective frequency of the single sample.
%
\begin{figure}[h!]
\center
  \psfrag{y}{$\!\!
y$}\psfrag{z}{$z$}\psfrag{a}{$\!\vec{J}_x$}\psfrag{b}{$\!\vec{J}_x$}
  a)\includegraphics[width=0.35\textwidth]{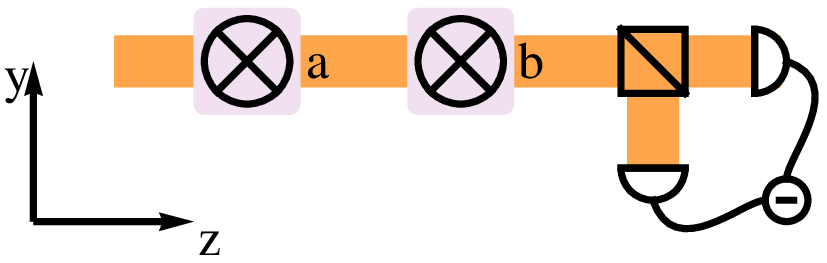}\quad b)\includegraphics[width=0.35\textwidth]{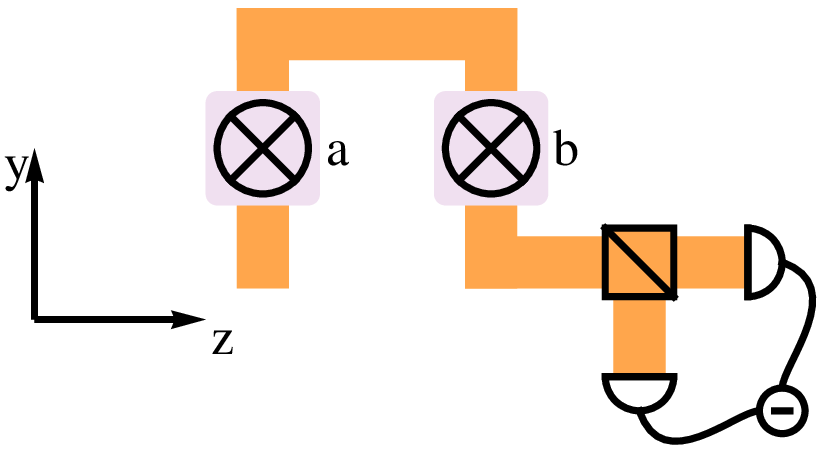}\\
  c)\includegraphics[width=0.35\textwidth]{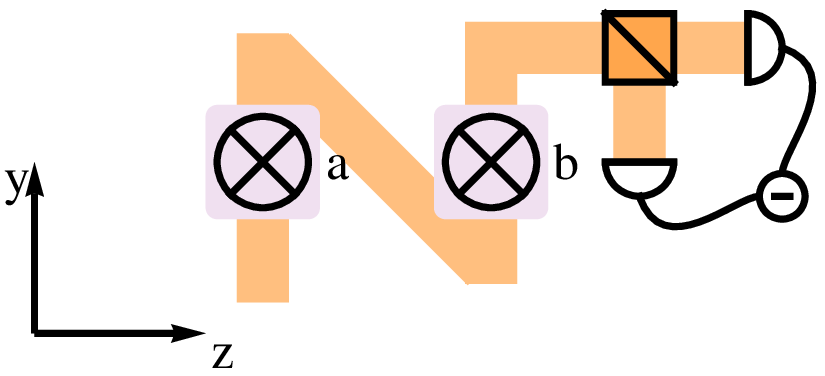}
  \caption{The sketch of the setups using the geometrical approach to generate
  and manipulate bipartite entanglement. The interaction between the light
  beam and atomic samples followed by the measurement introduces squeezing in
  a) $\hat{p}_{A,1}+\hat{p}_{A,2}$ b) $\hat{x}_{A,1}-\hat{x}_{A,2}$ and c)
$\hat{x}_{A,1}+\hat{x}_{A,2}$.}\label{fig:bipJulia}
\end{figure}

\noindent The QND Hamiltonain (\ref{hamiltonian}) for the many-mode setup
reduces to
\begin{equation}\label{hamiltonian_manymode}
\hat{H}_{\mathrm{int}}^{\mathrm{eff}}(\alpha)=-\sum_i \kappa_j
\hat{p}_{L} (\hat{p}_{A,i} \cos{\alpha_i}+\hat{x}_{A,i}
\sin{\alpha_i}).
\end{equation}
where $\alpha_{i}$ refers to the incident angle at which the light impinges the
atomic sample $i$. The corresponding symplectic matrix describing this
interaction is given by:
\begin{equation}\label{eqn:intsymplectic}
S_{\rm{int}}=\left(\begin{array}{ccc|c}
 &&              & G_1 \\
  &\hat{1}_{2N}^{A}& & \vdots \\
 &&              & G_N \\
 \hline
 M_1 & \cdots & M_N & \id_{2}^{L}
 \end{array}\right),
\end{equation}
with
\begin{equation}\label{eqn:G}
G_i=\left(\begin{array}{cc}
 -\kappa_i \sin \alpha_i & 0\\
 -\kappa_i \cos \alpha_i & 0
 \end{array}\right),\mbox{\hspace{1cm}}
M_j=\left(\begin{array}{cc}
  0 & 0 \\
 -\kappa_j \cos \alpha_j & \kappa_j \sin \alpha_j
 \end{array}\right).
\end{equation}
For the two mode atomic states, the covariance matrix of the atomic samples
after the interaction (setup in figure \ref{fig:bipJulia}b) can be
straightforwardly calculated from equation (\ref{symplectic2})
\begin{equation}
\gamma_{\rm{out}}=\left(
\begin{array}{cccc|cc}
 n_1+\kappa ^2 & 0 & \kappa ^2 & 0 & 0 & \kappa  \\
 0 & n_1 & 0 & 0 & n_1 \kappa  & 0 \\
 \kappa ^2 & 0 & n_2+\kappa ^2 & 0 & 0 & \kappa  \\
 0 & 0 & 0 & n_2 & n_2 \kappa  & 0 \\
 \hline
 0 & n_1 \kappa  & 0 & n_2 \kappa  & 1+n_1 \kappa ^2+n_2 \kappa ^2 & 0 \\
 \kappa  & 0 & \kappa  & 0 & 0 & 1
\end{array}
\right).
\end{equation}
\noindent Due to the atom-light interaction, both atomic modes are entangled
with light, however the reduced state of the two ensembles is separable as one
can easily check by applying the PPT criterion to the covariance matrix in the
upper-left block. Entanglement between atomic samples is not produced until one
measures a quadrature of light. Assuming the measurement outcome on
$\hat{x}_{L}$ to be $\tilde{x}_{L,1}$, the covariance matrix describing the
final state of the samples is given by [see (\ref{cm_measurement1}) and
(\ref{cm_measurement2})]
\begin{eqnarray}\label{cm_epr}
\gamma_{\rm{fin}}&=&\left(
\begin{array}{cccc}
 n_1+\kappa ^2 & 0 & \kappa ^2 & 0 \\
 0 & \frac{n_1 n_2 \kappa ^2+n_1}{\left(n_1+n_2\right) \kappa ^2+1} & 0 &
-\frac{n_1 n_2 \kappa ^2}{\left(n_1+n_2\right) \kappa ^2+1} \\
 \kappa ^2 & 0 & n_2+\kappa ^2 & 0 \\
 0 & -\frac{n_1 n_2 \kappa ^2}{\left(n_1+n_2\right) \kappa ^2+1} & 0 & \frac{n_1
n_2 \kappa ^2+n_2}{\left(n_1+n_2\right) \kappa ^2+1}
\end{array}
\right),
\end{eqnarray}
and the displacement of the final state is
\begin{equation}\label{displ_epr}
d_{\mathrm{fin}}=\left(0,-\frac{\tilde{x}_{L,1} \kappa n_1}{(n_1+n_2) \kappa
^2+1},0,-\frac{\tilde{x}_{L,1} \kappa n_1}{(n_1+n_2) \kappa ^2+1}\right).
\end{equation}
For what follows it is important to notice that the covariance matrix is
independent of the measurement outcome, but the latter is clearly present in the
displacement vector of the atomic modes. To check the entanglement between the
atomic samples after the light has been measured we use the separability
criterion based on the variances of the two commuting
operators\cite{Duan2000_sepCV}, which states that for any separable state the
total variances fulfill
\begin{equation}\label{crit}
\varsb{(|\lambda|\hat{p}_{A,1}+\frac{1}{\lambda}\hat{p}_{A,2})}+
\varsb{(|\lambda|\hat{x}_{A,1}-\frac{1}{\lambda}\hat{x}_{A,2})}\geq 2\hbar.
\end{equation}
This is a sufficient but not necessary condition for separability.

We restrict our analysis to a single inequality involving the
collective observables with $\lambda=1$ in (\ref{crit}) since it is the
one applicable experimentally \cite{Julsgaard2001_entanglement}. The way to
measure such combination of variances has been described in detail in
\cite{Stasinska2009_meso}.

\begin{figure}[h!]
 a)\includegraphics[width=0.4\textwidth]{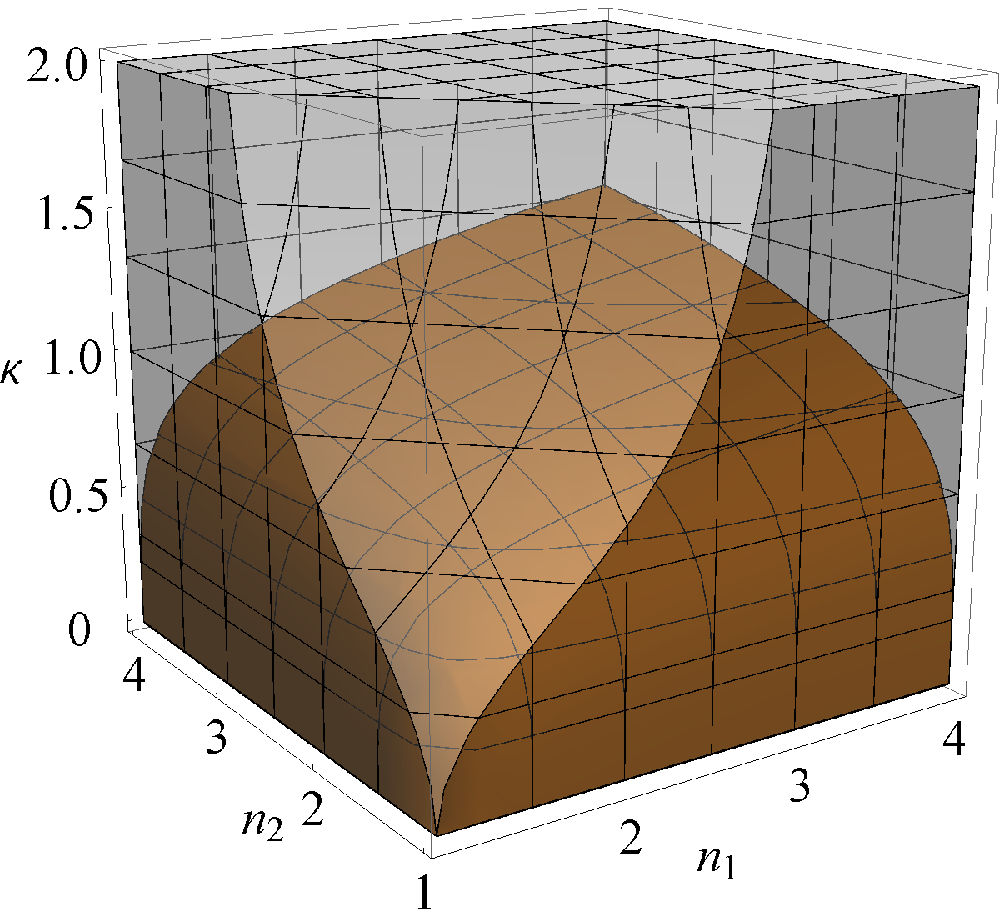}\quad b) \includegraphics[width=0.4\textwidth]{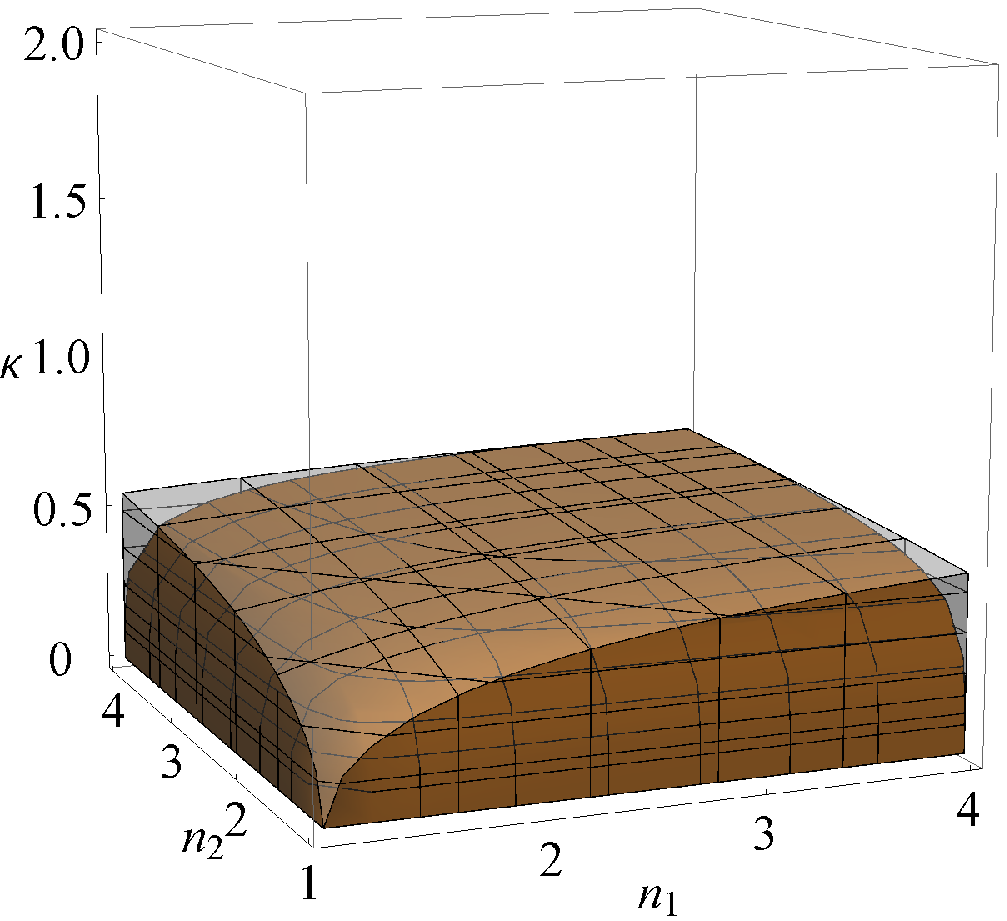}\\
  \caption{Comparison of the sets of parameters $\kappa, n_1, n_2$ for which the
state of two ensembles remains separable (brown inner region), with the ranges
of parameters for which it is not detected by the spin variance inequality
(\ref{crit}) with $\lambda=1$ (gray outer region). In figure a) we consider the
state produced in the setup in figure \ref{fig:bipJulia}a, whereas in figures
b) the one produced in two steps schematically depicted in
figures \ref{fig:bipJulia}a and \ref{fig:bipJulia}b.}\label{ppt_thermal}
\end{figure}

Extracting the variances from the elements of the final covariance
matrix (\ref{cm_epr}) we obtain
\begin{eqnarray}
\frac{1}{\hbar}\varsb{(\hat{p}_{A,1}+\hat{p}_{A,2})}&=&\frac{1}{2} (\gamma_{
\mathrm{fin},22}+\gamma_{\mathrm{fin},44}+2
\gamma_{\mathrm{fin},24})=\frac{n_1+n_2}{2 \left(n_1+n_2\right) \kappa
^2+2},\nonumber\\
\frac{1}{\hbar}\varsb{(\hat{x}_{A,1}-\hat{x}_{A,2})}&=&\frac{1}{2} (\gamma_{
\mathrm{fin},11}+\gamma_{\mathrm{fin},33}-2
\gamma_{\mathrm{fin},13})=\frac{1}{2} \left(n_1+n_2\right).
\end{eqnarray}
The substitution of the above expressions in equation (\ref{crit}) leads to the
violation of the separability criterion for the values of $n_1,n_2$, and
$\kappa$ fulfilling
\begin{eqnarray}\label{var_thermal}
\kappa^2>\frac{2 (n_1+n_2-2)}{\left(4-n_1-n_2\right)
\left(n_1+n_2\right)},\quad n_1+n_2<4\\
\kappa^2<\frac{2 (n_1+n_2-2)}{\left(4-n_1-n_2\right)
\left(n_1+n_2\right)},\quad n_1+n_2>4
\end{eqnarray}
One immediately notes that for $n_1+n_2>4$, the inequality can never be
violated, since the right-hand side of the inequality becomes negative. In
Figure \ref{ppt_thermal} we compare the complement of the set defined by
(\ref{var_thermal}) (states that are not detected) with the set of separable
(PPT) states. This shows that the variance inequality does not detect all
entangled states. This is due to the fact that only one combination of
variables, i.e., $\hat p_{A,1}+\hat p_{A,2}$ is squeezed. In this case the
thermal fluctuations still present in $\hat x_{A,1}-\hat x_{A,2}$  does not
allow the sum in equation (\ref{crit}) to violate the bound.

In the case of initially pure states an arbitrary coupling produces entanglement
\cite{Stasinska2009_meso}. Here we see that for initially thermal states this is
not the case and the generation of entanglement requires a stronger coupling
with light.

In order to increase the amount of entanglement and improve its detectability
through the variance inequality criterion, a sequence of steps displayed in
Figure \ref{fig:bipJulia}a and \ref{fig:bipJulia}b is required. These introduce
squeezing in two commuting combinations of quadratures. Calculation similar to
those from the previous paragraph lead to
\begin{eqnarray}
\frac{1}{\hbar}\varsb{(\hat{p}_{A,1}+\hat{p}_{A,2})}=\frac{1}{\hbar}\varsb{(\hat
{x}_{A,1}-\hat{x}_{A,2})}=\frac{n_1+n_2}{2
\left(n_1+n_2\right) \kappa ^2+2},
\end{eqnarray}
and the violation of the variance's inequality (\ref{crit}) for $\lambda=1$ is
now obtained for a much lower coupling:
\begin{equation}
\kappa^2>\frac{n_1+n_2-2}{2 n_1+2 n_2}.
\end{equation}
The set of states detected by the spin variance inequality is compared in figure
\ref{ppt_thermal}b to the set of separable (equivalently PPT) states. Again the
spin variance inequality does not detect all entangled states, however, now is
much more efficient than in the previous case since the fluctuations in both
combinations of variables were suppressed.

Interesting enough such measurement induced entanglement between two atomic samples
can be deleted by exploiting the squeezing and anti-squeezing effects produced
by the laser beams for pure state entanglement\cite{Stasinska2009_meso}. Here
we demonstrate that this is also true when the initial states are thermal,
however, the procedure has to be slightly modified. Let us consider the state
represented by the covariance matrix (\ref{cm_epr}). Interaction with a second
light beam impinging on each atomic sample at $\alpha_1=\alpha_2=\pi/2$ (as
depicted in Figure \ref{fig:bipJulia}c) followed by the light measurement will
erase the entanglement produced by the interaction with the first light beam if
the properties of light are appropriately adjusted. To this aim, it is
sufficient to impose a second light beam characterized by the covariance matrix
\begin{equation}\label{eq:beam_gen}
\gamma_L=\left(%
\begin{array}{cc}
  (\triangle x_1)^2 & 0 \\
  0 & (\triangle p_1)^2 \\
\end{array}%
\right),
\end{equation}
with
\begin{eqnarray}
&&(\triangle x_1)^2=\kappa ^2 n_1+\kappa ^2 n_2+n_1 n_2\nonumber\\
&&(\triangle p_1)^2=\frac{n_1 n_2}{\kappa ^2 n_1+\kappa ^2 n_2+1}.
\end{eqnarray}
and interaction coupling $\kappa$. The covariance matrix of
the final state is given by:

\begin{equation}\label{eq:cm_erased}
\gamma_{\mathrm{er}}=\left(
\begin{array}{cccc}
 \frac{\kappa ^2 n_2+n_1 \left(\kappa ^2+n_2\right)}{2 \kappa ^2+n_2} & 0 &
   0 & 0 \\
 0 & \frac{n_1 \left(2 \kappa ^2 n_2+1\right)}{\kappa ^2
   \left(n_1+n_2\right)+1} & 0 & 0 \\
 0 & 0 & \frac{\kappa ^2 n_2+n_1 \left(\kappa ^2+n_2\right)}{2 \kappa
   ^2+n_1} & 0 \\
 0 & 0 & 0 & \frac{n_2 \left(2 \kappa ^2 n_1+1\right)}{\kappa ^2
   \left(n_1+n_2\right)+1}
\end{array}
\right),
\end{equation}
and the corresponding displacement is:
\begin{equation}
d_{\mathrm{er}}=\left\{-\frac{\tilde{x}_{L,2} \kappa}{2 \kappa
^2+n_2},-\frac{\tilde{x}_{L,1} \kappa n_1}{\kappa
   ^2 \left(n_1+n_2\right)+1},-\frac{\tilde{x}_{L,2} \kappa}{2 \kappa
   ^2+n_1},-\frac{\tilde{x}_{L,1} \kappa  n_1}{\kappa ^2
   \left(n_1+n_2\right)+1}\right\},
\end{equation}
where $\tilde{x}_{L,2}$ is the measurement outcome. Note that the state
(\ref{eq:cm_erased}) is separable. It will be identical to the initial
one, however, only if $n_2=n_1$.

\subsection{ Multipartite bound
entanglement}\label{subs:multipartite}

We move now to the truly multipartite entanglement and focus on the simplest
case of the 3 atomic samples. We aim at analyzing bound entanglement, which
exist only in the mixed state case. The classification of tripartite
entanglement in CV was given by Giedke and coworkers in
\cite{Giedke2001_Gauss3mode}, where five classes of states were distinguished.
Following their classification, inseparable states with respect to every
bipartite splitting are denoted as class 1. States which are biseparable
with respect to one (and only one) bipartition belong to class 2. States that
are biseparable with respect to two or three bipartitions, but still entangled,
belong to class 3 and class 4, respectively. These two classes of states are
bound entangled. Finally, class 5 are the separable states.

Continuous-variable cluster-like states can be a universal resource for optical
quantum
computation\cite{Nielsen2004_cluster,Ohliger2010_limitations,
Aolita2011_thermalCVcluster}. They are defined by analogy with the discrete
cluster states generated via Ising interactions between qubits
\cite{Aolita2011_thermalCVcluster}.  One can associate the modes of the $N$-mode
(CV) system with the vertices of a graph $G$. And the cluster corresponds to a
connected graph\cite{Briegel2001_cluster}. For continuous variables, cluster states
are defined only asymptotically as those with infinite squeezing in the variables
\begin{equation}\label{eq:cluster_def}
\hat{p}_{a}-\sum_{b\in N_{a}}\hat{x}_{b}
\end{equation}
for all the modes belonging to the graph, where $N_{a}$ denotes the set of
neighbors of vertex $a$. Cluster-like states are defined when the squeezing
is finite.

It is possible to create such states with atomic ensembles in the proper
geometrical setup \cite{Stasinska2009_meso} taking the samples in the initial
vacuum state. The protocol consists of interaction with light passing through
the samples at specified angles (see Table \ref{tab:cluster}) plus a homodyne
detection of light. Here we check the properties of the states obtained through
the same procedure, however for initially thermal samples, i.e.,
$\gamma_{in}^{(0)}=n \id_{6} \oplus \id_2$. We demonstrate that in such setup
bound entangled states are produced for certain choices of temperature and
coupling. The presence of undistillable entanglement in thermal
finite-dimensional systems was considered in \cite{Cavalcanti2010a}.

We analyze the two cluster states, the linear one and the triangular one, focusing
on how the temperature destroys the genuine tripartite entanglement. To
discriminate class 1, 2 and 3, the PPT criterion is enough. To discriminate
between class 4 and 5 we will use the operational necessary and sufficient
criterion for full separability given in \cite{Giedke2001_Gauss3mode}.

\begin{table}[!h]
\begin{center}
 \begin{tabular}{c|ccc|ccc|cccc}
     & \multicolumn{3}{|c|}{Linear} & \multicolumn{3}{|c}{Triangular} & \multicolumn{4}{|c}{Smolin state}\\
        \hline
   beam &  $\alpha_1$ & $\alpha_2$      & $\alpha_3$ & $\alpha_1$ & $\alpha_2$ & $\alpha_3$ & $\alpha_1$ & $\alpha_2$ & $\alpha_3$ & $\alpha_4$\\
   \hline
   1 & 0 & $\frac{\pi}{2}$& -    & $0$ & $\frac{\pi}{2}$ & $\frac{\pi}{2}$ & $0$ & $0$ & $\pi$ & $\pi$\\
   2 & $\frac{\pi}{2}$& 0 &$\frac{\pi}{2}$ & $\frac{\pi}{2}$ &  $0$ & $\frac{\pi}{2}$ & $\frac{\pi}{2}$ & $\frac{\pi}{2}$ & $-\frac{\pi}{2}$ & $-\frac{\pi}{2}$\\
   3 & - &$\frac{\pi}{2}$ & $0$ & $\frac{\pi}{2}$ & $\frac{\pi}{2}$ & $0$ & - & - & - & -\\
 \end{tabular}
\caption{Parameters for tripartite cluster states and the four-partite Smolin
state.}\label{tab:cluster}
\end{center}
\end{table}

\begin{figure}[!t]
\begin{center}
(a)\includegraphics[trim=80 50 20 10,
width=0.3\textwidth,angle=-90]{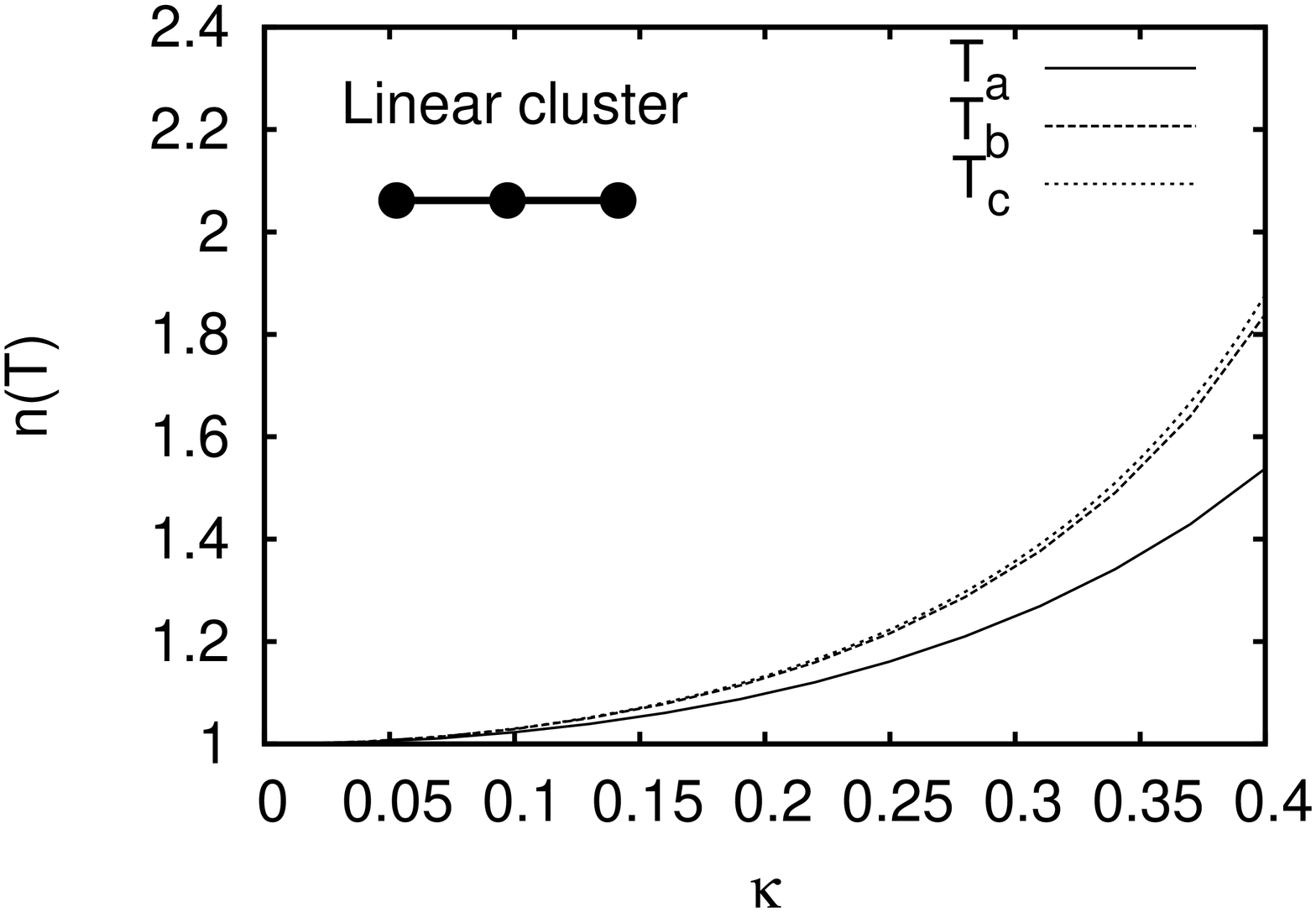}(b)\includegraphics[trim=80
50 20 10,width=0.3\textwidth,angle=-90]{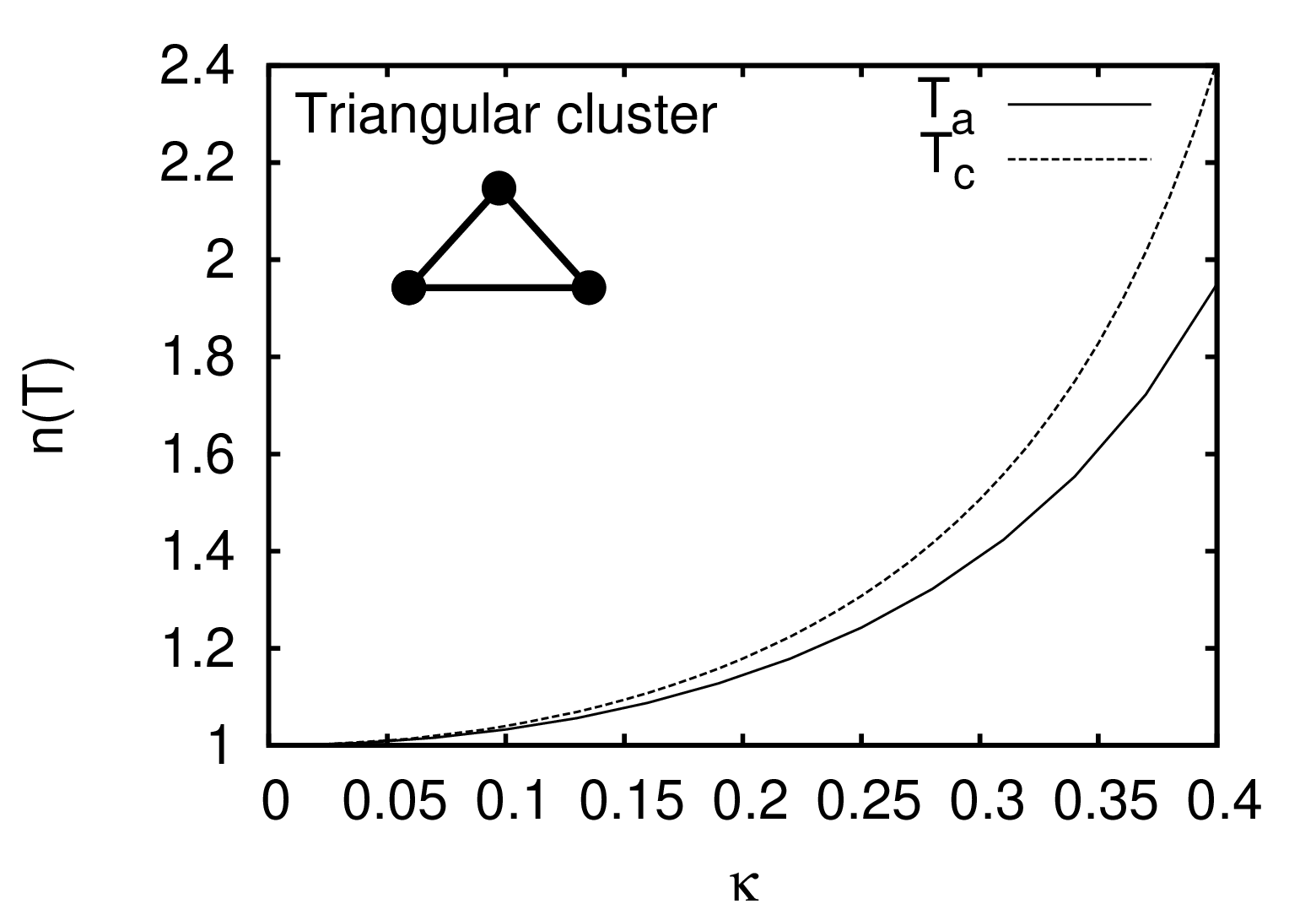}
\caption{Range of parameters for (a) the linear and (b) the triangular cluster
state for which it belongs to different entanglement classes (see explanation
it the text).} \label{fig:cluster}
\end{center}
\end{figure}

The results we obtain are summarize in Figure \ref{fig:cluster}
where we depict as a function of the temperature $T$ through
$n(T)=1/\tanh\left(\omega/2T\right)$ and coupling $\kappa$, the
different entanglement types created between the atomic samples.

For the linear cluster state (figure \ref{fig:cluster}a), for a fixed value
of the parameter $\kappa$, the state is NPT with respect to all the three cuts
in the interval $0<T<T_a$, meaning that it belongs to class 1. For $T_a<T<T_b$
the state is PPT with respect to the bipartition $2-(1,3)$ and NPT with respect
to others. Hence, in it belongs to bound entangled class 3. For $T>T_b$ the
state becomes PPT with respect to all the cuts. Using the separability criterion
found in \cite{Giedke2001_Gauss3mode}, we find that within the range $T_b<T<T_c$
the state is class-4, while for $T>T_c$ it becomes fully separable.

Figure \ref{fig:cluster}b corresponds to the different entanglement classes for
a triangular cluster state. In this case, because of the permutational
invariance, class 3 is never recovered. Nevertheless a class-4 region still
appears between the genuine tripartite and the fully separable states.

\subsection{Multipartite unlockable bound entanglement}\label{subs:smolin}
Here we propose to use the Gaussian randomness that is inherently present in the
system due to uncertainty of the outcome of the light measurement. We will show
that it is essential to generate the bound entangled Smolin state.

The Smolin state introduced in~\cite{Smolin2001_be} using qubits
is an interesting example of an undistillable multipartite entangled state that
can be unlocked if several parties perform a collective measurement and send the
outcome to the remaining parties. The Smolin state if form by an equal-weight
mixture of products of the four Bell states $\psi_i (i=1,\ldots,4)$
\begin{equation}\label{eq:Smolin_state}
\varrho_{\mathrm{Smolin}}=\sum_{i=1}^{4}
\proj{\psi_i}_{AB}\ot\proj{\psi_i}_{CD}.
\end{equation}
One sees that parties $AB$ and $CD$ share the same  Bell state but none of the
pairs know which one it is. Such state is clearly separable with respect to
partition $AB|CD$. Moreover, since it is permutationally invariant, the above
statement also holds for every two pairs, not only for $AB$ and $CD$. Hence,
according to the definition, the state is undistillable with respect to
arbitrary two parties, since there always exists a separable bipartition
dividing them.

The fact that arbitrary two pairs share the same unknown Bell state makes the
entanglement present in the Smolin state unlockable. Imagine the situation in which
two of the parties, say $C$ and $D$, meet in one laboratory and perform
collectively a measurement in the Bell basis, identifying thus the state they
possess. This measurement projects the state shared by $AB$ on the one
corresponding to the outcome of the parties $CD$. Hence, if $A$ and $B$, still
being in separate laboratories, obtain (by classical communication) the
information about the outcome of $CD$, they can use their maximally entangled
state. We emphasize that the otherwise unknown Bell state is useless.

The Smolin state possesses many other interesting features. Apart from being
unlockable, its entanglement can be activated \cite{Shor2003_superact} through
a cooperation of five parties sharing two copies of the state. Despite being
bound entangled, this state can be used in the remote quantum information
concentration protocol \cite{Murao2001} and maximally violates a Bell inequality
being at the same time useless for the secret key distillation
\cite{Augusiak2006_Smolin}.

\begin{figure}[!t]
\begin{center}
\includegraphics[trim=10 40 10 30,
width=0.65\textwidth]{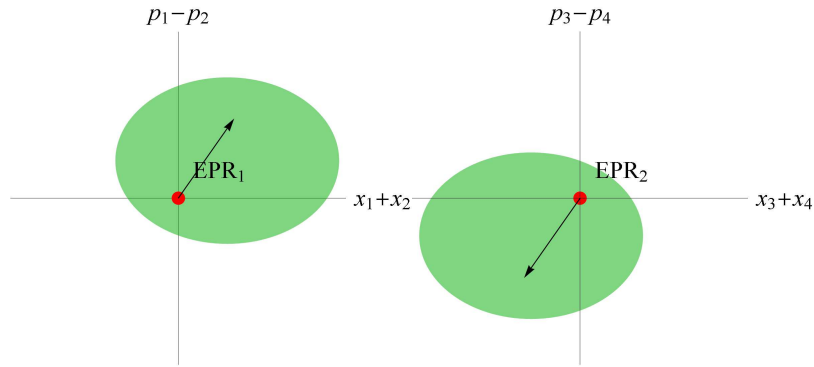}%
\caption{Schematically depicted Smolin state in the phase space. The red dots
correspond to the EPR-like states, the green ellipses represent the random
Gaussian displacement that depends on the properties of light passing through
the samples.} \label{fig:smolin}
\end{center}
\end{figure}

Recently, Zhang has generalized the discrete Smolin state to the
Gaussian CV scenario using the mathematical formalism of stabilizers
\cite{Zhang2011_Smolin}. It turns out that the state proposed by Zhang can
indeed be understood as some mixture of products of the continuous variables
EPR-like states in this way. Let us consider two EPR(-like) states with known
reference displacement and randomly displace both of them in the opposite
\emph{unknown} directions in the phase space, as schematically depicted in
figure \ref{fig:smolin}. The only constraint imposed on the random displacement
is that it has a Gaussian distribution. In this way we obtain a Gaussian mixture
of two displaced EPR pairs.

The properties of the CV Smolin state are slightly different from those of the
qubit state. Firstly and most importantly, we  loose the permutational
invariance \cite{Zhang2011_Smolin}. Secondly, since the maximally entangled
EPR-state corresponds to infinite squeezing in the CV and is not physical, we
will always have to deal with "EPR-like" states and therefore it is necessary to
analyze the effect of the finite squeezing on the properties of the final
Smolin-like state. Both of the above constraints force the Smolin-like state to
be undistillable only with respect to chosen parties and within a particular
range of the squeezing parameter.

Let us demonstrate how the mathematical setup generating the Smolin-like
state proposed in \cite{Zhang2011_Smolin} greatly simplifies in the
atomic-ensemble realization. We characterize each step of the procedure using
the covariance matrix and the displacement vector. Finally, using the symplectic
formalism, we demonstrate the unlocking protocol.

From now on we will distinguish the parties (modes) with Latin numbers. As
demonstrated by Zhang, the Smolin-like state can be generated in two steps:

\noindent (i) generate two EPR-like pairs with reduced fluctuations (squeezing)
in $\hat x_1+\hat x_2$ and $\hat p_1-\hat p_2$ ($\hat x_3+\hat x_4$ and $\hat
p_3-\hat p_4$ for modes $3$ and $4$) and known displacement\\

\noindent (ii) displace the EPR pairs by random Gaussian variables $\lambda_1$
and $\lambda_2$ such that the quadratures transform in the following way:
\begin{equation}\label{smolin_quadr1}
\hat{x}_1+\hat{x}_2+\hat{x}_3+\hat{x}_4 \rightarrow
(\hat{x}_1-\lambda_1)+(\hat{x}_2-\lambda_1)+(\hat{x}_3+\lambda_1)+(\hat{x}_4+\lambda_1),
\end{equation}
and
\begin{equation}\label{smolin_quadr2}
\hat{p}_1-\hat{p}_2+\hat{p}_3-\hat{p}_4 \rightarrow
(\hat{p}_1+\lambda_2)-(\hat{p}_2-\lambda_2)+(\hat{p}_3-\lambda_2)-(\hat{p}_4+\lambda_2).
\end{equation}
Note that indeed the squeezed combinations in modes $1-2$ and $3-4$ are displaced
in the opposite directions. Implementation of each of these steps is experimentally
feasible in the geometrical setup of atomic ensembles. Step (i) concerns the
generation of EPR-like states with known displacement which is easily achieved
(see also \cite{Stasinska2009_meso}). Step (ii) concerns the generation of a
random Gaussian variables in the displacements. In our geometrical setup this
can be achieved by shining two light beams at some given angles as summarized in
table \ref{tab:cluster}. A straightforward application shows that the angles
reported in the above table modify the the quadratures of atomic spin ensembles
in the following way:
\begin{equation}\label{smolin_quadr_our1}
\hat{x}_1+\hat{x}_2+\hat{x}_3+\hat{x}_4 \rightarrow
(\hat{x}_1-\kappa \hat{p}_{L,1})+(\hat{x}_2-\kappa
\hat{p}_{L,1})+(\hat{x}_3+\kappa \hat{p}_{L,1})+(\hat{x}_4+\kappa
\hat{p}_{L,1})
\end{equation}
and
\begin{equation}\label{smolin_quadr_our2}
\hat{p}_1-\hat{p}_2+\hat{p}_3-\hat{p}_4 \rightarrow
(\hat{p}_1+\kappa \hat{p}_{L,2})-(\hat{p}_2-\kappa
\hat{p}_{L,2})+(\hat{p}_3-\kappa \hat{p}_{L,2})-(\hat{p}_4+\kappa
\hat{p}_{L,2}).
\end{equation}
Comparing now equations (\ref{smolin_quadr1}) and (\ref{smolin_quadr2}) to
equations (\ref{smolin_quadr_our1}) and (\ref{smolin_quadr_our2}) respectively,
one observes that our setup reproduces precisely the required displacements.
Now, if the output light is \emph{not measured}, $\kappa \hat{p}_{L,1}$ and
$\kappa \hat{p}_{L,2}$ remain \emph{unknown} random Gaussian variables with
expectation value and variance that can be easily adjusted by choosing the input
light with specific properties (vacuum squeezed, thermal etc.). From now on we
will denote this random variables by $\bar{p}_{L,1},\bar{p}_{L,2}$.

We switch to the covariance matrix formalism that allows us to get a better
insight into the process and the properties of the final state. We begin
directly with two $EPR$ pairs with zero reference displacement and characterized
by the squeezing parameter $r$, and a general light mode characterized by the
variances: $\var{\hat{x}_{L,1}},\var{\hat{p}_{L,1}}$ and zero displacement.
The initial state of the setup is described by:
\begin{equation}
\gamma_{1}^{(\mathrm{in})}
=\gamma_{\mathrm{EPR_1}}\oplus\gamma_{\mathrm{EPR_2}}\oplus
\left(\begin{array}{cc}
\var{\hat{x}_{L,1}} & 0\\
0 & \var{\hat{p}_{L,1}}\\
\end{array}\right).
\end{equation}
The interaction between the atomic ensembles and the light leads to the
following CM for atoms
\begin{eqnarray}
\gamma_{1}^{(\mathrm{out})} =\left(
\begin{array}{cccccccc}
 a_1 & 0 & c_1 & 0 & e_1 & 0 & e_1 & 0 \\
 0 & b & 0 & d & 0 & 0 & 0 & 0 \\
 c_1 & 0 & a_1 & 0 & e_1 & 0 & e_1 & 0 \\
 0 & d & 0 & b & 0 & 0 & 0 & 0 \\
 e_1 & 0 & e_1 & 0 & a_1 & 0 & c_1 & 0 \\
 0 & 0 & 0 & 0 & 0 & b & 0 & d \\
 e_1 & 0 & e_1 & 0 & c_1 & 0 & a_1 & 0 \\
 0 & 0 & 0 & 0 & 0 & d & 0 & b
\end{array}
\right)\\
\mathrm{with}\nonumber \\
a_1=\kappa ^2\var{\hat{p}_{L,1}}+\ch 2 r, \quad b=\ch
2r\quad c_1=\kappa ^2 \var{\hat{p}_{L,1}}-\sh 2 r,\nonumber\\
d=\kappa^2\var{\hat{p}_{L,1}}-\ch 2r \quad e_1=-\kappa ^2
\var{\hat{p}_{L,1}}. \nonumber
\end{eqnarray}

Interaction with a second mode of light characterized by variances
$\var{\hat{x}_{L,2}}, \var{\hat{p}_{L,2}}$, leads to the CM:
\begin{eqnarray}
\gamma_{2}^{(\mathrm{out})} = \left(
\begin{array}{cccccccc}
a_1 & 0 & c_1 & 0 & e_1 & 0 & e_1 & 0 \\
0 & a_2 & 0 & -c_2 & 0 & e_2 & 0 & -e_2 \\
c_1 & 0 & a_1 & 0 & e_1 & 0 & e_1 & 0 \\
0 & -c_2 & 0 & a_2 & 0 & -e_2 & 0 & e_2 \\
e_1 & 0 & e_1 & 0 & a_1 & 0 & c_1 & 0 \\
0 & e_2 & 0 & -e_2 & 0 & a_2 & 0 & -c_2 \\
e_1 & 0 & e_1 & 0 & c_1 & 0 & a_1 & 0 \\
0 & -e_2 & 0 & e_2 & 0 & -c_2 & 0 & a_2
\end{array}
\right)\\
a_1=\kappa ^2\var{\hat{p}_{L,1}}+\ch 2 r, \quad a_2=\kappa
^2\var{\hat{p}_{L,2}}+\ch 2 r, \nonumber\\
c_1=\kappa ^2 \var{\hat{p}_{L,1}}-\sh 2r,\quad c_2=\kappa ^2
\var{\hat{p}_{L,2}}-\sh 2 r,\nonumber\\
e_1=-\kappa ^2\var{\hat{p}_{L,1}},\quad e_2=-\kappa ^2
\var{\hat{p}_{L,2}}.\nonumber
\end{eqnarray}
and the displacement of the final state is given by:
\begin{equation}
d_2^{(\mathrm{out})}=\left(-\kappa \bar{p}_{L,1},\kappa
\bar{p}_{L,2},-\kappa \bar{p}_{L,1},-\kappa \bar{p}_{L,2},\kappa
\bar{p}_{L,1},-\kappa \bar{p}_{L,2},\kappa \bar{p}_{L,1},\kappa
\bar{p}_{L,2}\right).
\end{equation}
We verify separability of the obtained state using the partial time reversal
criterion and see that the state:\\
(i) it is always $PPT$ with respect to partition $12|34$,\\
(ii) it is always $NPT$ with respect to partition $13|24$,\\
(iii) it is $PPT$ with respect to partition $14|23$ only if
\begin{equation}
\kappa^2 \var{\hat p_{L,1}}\geq \frac{1}{4} \left(e^{2 r}-e^{-2
r}\right)\; \mathrm{or}\; \kappa^2 \var{\hat p_{L,2}}\geq
\frac{1}{4} \left(e^{2 r}-e^{-2 r}\right);
\end{equation}
(iv) all the partitions one vs. three modes are always $NPT$, in agreement with
the results obtained in \cite{Zhang2011_Smolin}. \\
An important thing to notice is that the state is unlockable bound entangled
only if the partition $14|23$ is $PPT$, since only in this case the entanglement
between modes $1$ and $2$ is bound and the unlocking procedure makes sense.

In order to show the unlockability of the obtained state, we write the CM and
displacement in the basis in which the EPR states are diagonal, i.e.,
$(1/\sqrt{2})\left(\hat x_1+\hat x_2,\ldots,\hat p_3-\hat p_4\right)$. For
simplicity we also assume that $\var{\hat x_{L,1}}=\var{\hat x_{L,2}}=\var{\hat
x_{L}}$ and $\var{\hat p_{L,1}}=\var{\hat p_{L,2}}=\var{\hat p_{L}}$ then
\begin{eqnarray}\label{eq:SmolinEPR}
\gamma_{\mathrm{S}}=\left(
\begin{array}{cccccccc}
 e^{- 2 r}+f & 0 & 0 & 0 & -f & 0 & 0 & 0 \\
 0 & e^{2 r}& 0 & 0 & 0 & 0 & 0 & 0 \\
 0 & 0 & e^{2 r} & 0 & 0 & 0 & 0 & 0 \\
 0 & 0 & 0 & e^{- 2 r}+f & 0 & 0 & 0 & -f \\
 -f & 0 & 0 & 0 & e^{- 2 r}+f & 0 & 0 & 0 \\
 0 & 0 & 0 & 0 & 0 & e^{2 r} & 0 & 0 \\
 0 & 0 & 0 & 0 & 0 & 0 & e^{2 r} & 0 \\
 0 & 0 & 0 & -f & 0 & 0 & 0 & e^{- 2 r}+f
\end{array}
\right)\\
f=2 \kappa^2 \var{\hat p_{L}}\nonumber
\end{eqnarray}

\begin{equation}
d_{\mathrm{S}}=\left\{-\sqrt{2} \kappa \bar{p}_{L,1},0,0,\sqrt{2}
\kappa \bar{p}_{L,2},\sqrt{2} \kappa \bar{p}_{L,1},0,0,-\sqrt{2}
\kappa
   \bar{p}_{L,2}\right\}
\end{equation}
Now inspection of $\gamma_s$ shows clearly that we have two correlated $EPR$
pairs displaced by  random vectors pointing in the opposite directions in the
phase space of the squeezed variables, exactly as depicted in figure
\ref{fig:smolin}. Since we do not know where in the phase space are the EPR
pairs, we cannot use them for quantum tasks. Therefore, by analogy to the
discrete case \cite{Smolin2001_be,Augusiak2006_Smolin}, unlocking of
entanglement is understood as learning which of the displaced $EPR$ pairs is
shared by pair $1-2$ by measuring the displacement of the other EPR pair shared
by $3-4$.

For completeness we also demonstrate all the steps of the unlocking protocol
with the atom-light interface using the symplectic formalism. The measurement of
$\hat{x}_3+\hat{x}_4$ and $\hat{p}_3-\hat{p}_4$ can be done with two probe
light beams with reduced fluctuations in $x_P$, i.e., $\var{x_P}<\hbar/2$ (QND
measurement) \cite{Sorensen1998_spectroscopy}. The probe beams pass through the
samples $3-4$ at the angles $\alpha_3^{(1)}=\alpha_4^{(1)}=\pi/2$ and
$\alpha_3^{(2)}=0,\alpha_4^{(2)}=\pi$, respectively, and are measured
afterwards. If the measurement outcomes are, respectively, $x_{+}$, $p_{-}$, the
final state of modes $1-2$ is characterized by the following CM and displacement
(still in the $EPR$ basis):
\begin{equation}\label{unlocked}
\hspace{-1.2cm}\gamma_{12}=\left(
\begin{array}{cccc}
\frac{f\left(2 \kappa ^2+e^{2 r} \var{\hat
x_P}\right)}{e^{2 r} \left(2
\kappa ^2 f+\var{\hat x_P}\right)+2 \kappa ^2}+e^{-2 r} & 0 & 0 & 0 \\
 0 & e^{2 r} & 0 & 0 \\
 0 & 0 & e^{2 r} & 0 \\
 0 & 0 & 0 & \frac{f\left(2 \kappa ^2+e^{2 r} \var{\hat
x_P}\right)}{e^{2 r} \left(2
\kappa ^2 f+\var{\hat x_P}\right)+2 \kappa ^2}+e^{-2 r}
\end{array}
\right),
\end{equation}
\begin{eqnarray}\label{eq:unlocked_cm}
\hspace{-1.2cm}d_{12}=\sqrt{2}\kappa \times \left[x_{+} g-
\bar{p}_{L,1},0,0,p_{-} g+\bar{p}_{L,2}\right],\\
\hspace{-1.2cm}g=\frac{f}{\var{\hat x_P}+2 \kappa^2( e^{-2
r}+f)}\nonumber
\end{eqnarray}
If the variables $\hat{x}_3+\hat{x}_4$ and $\hat{p}_3-\hat{p}_4$
are strongly enough squeezed, the obtained outcome approximates
well the displacement of $\hat{x}_1+\hat{x}_2$ and $\hat{p}_1-\hat{p}_2$.
Therefore, we may write $2\kappa \bar{p}_{L,1}=-x_{+}$ and $2\kappa
\bar{p}_{L,2}=-p_{-}$. In this way the resulting displacement is a function of
the known parameters
\begin{equation}\label{eq:unlocked_disp}
\hspace{-1.2cm}\tilde{d}_{12}=\sqrt{2} \left[x_{+} \left(\kappa
g+\frac{1}{2}\right),0,0,p_{-} \left(\kappa
g-\frac{1}{2}\right)\right].
\end{equation}

\begin{figure}
  \centering
  \includegraphics[width=0.5\textwidth]{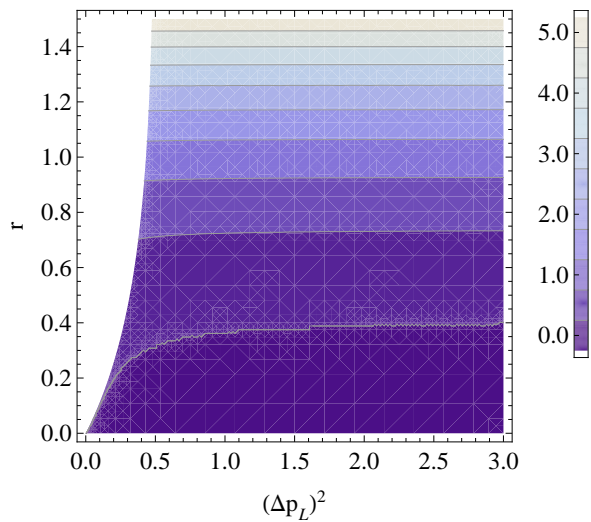}\\
  \caption{Negativity of the unlocked state given by equation (\ref{eq:unlocked_cm})
  as a function of the squeezing parameter $r$ and variance of the random
displacement $\var{\hat p_L}$. All the other parameters are expressed in terms
of $r$ (see explanation in the text).}\label{fig:negativity}
\end{figure}

In comparison to the initial EPR state, the covariance matrix (\ref{unlocked})
has an additional positive term in diagonal elements, which reads
\begin{equation}
\hspace{-1.2cm}\delta=\frac{f\left(2 \kappa ^2+e^{2 r} \var{\hat
x_P}\right)}{e^{2 r} \left(2 \kappa ^2 f+\var{\hat x_P}\right)+2
\kappa ^2}.
\end{equation}
It increases the variance of the squeezed variables and therefore makes the
state mixed and less entangled. In order to quantify the entanglement in the
final unlocked state, we assume that the coupling constants $\kappa$ used in the
preparation of the EPR states and the ones used for generation and unlocking of
the Smolin state are equal and we express them in terms of the squeezing
parameter $r$
\begin{equation}
\kappa^2=(1+\e^{2r})/2.
\end{equation}
Further we assume that the squeezing of the probe light beam is the same as the
squeezing of initial EPR pairs, i.e., $\var{\hat x_P}=e^{-2r}$. In this way, the
properties of the final state depend only on two parameters: $r$ and $\var{\hat
p_L}$. In figure \ref{fig:negativity} we plot the negativity of the unlocked
state, as a function of these parameters. For the admissible values of $r$, the
negativity of the final state is roughly a half of the value for the initial EPR
pair.

\section{Conclusions}\label{sec:summary}
Our results can be summarize us follows.  We have shown that, unlike in the pure
state case, if the atomic ensembles are initially in a thermal state the
entangling procedure only succeeds for certain choices of the coupling
parameter. In this way we have provided bounds on the experimental parameters
leading to mixed state entanglement. Further we have shown that tripartite
continuous cluster-like states are robust against initial thermal noise present
in the atomic ensembles, meaning that there exists a finite range of
temperatures for which the final state remains genuine tripartite entangled. We
have also demonstrate that with increasing temperature, the states becomes bound
entangled and finally fully separable.

Finally, we have also shown that a mixed state of atomic ensembles can be
produced from an initial pure state if the light beam is not measured after the
interaction. In this case, the interaction with the light beam mixes all the
possible outcomes with a Gaussian weight. Using this procedure, starting from
two pure bipartite entangled states, it is possible to produce the so-called
Smolin state which is an example of an unlockable bound entangled state. It
should be emphasized that, since this procedure does not involve the final
projective measurement, it cannot produce extra entanglement. Its only effect is
introducing randomness in the system, making the entanglement bound.

Summarizing, our analysis shows that atomic ensembles offer a versatile toolbox
for generation and manipulation of multipartite entanglement both in  the pure
and mixed state cases. By exploiting further the geometrical setting introduced
in \cite{Stasinska2009_meso}  we have been able to assess the robustness of the
entangling schemes to set guidelines for future experiments, both in the
bipartite as well as in the multipartite case. Moreover, we have demonstrated
that the very same setting provides experimentally feasible schemes for
generation of bound entanglement, broadening the applicability of atomic
ensembles for quantum information studies. In particular, our results provide,
to the best of our knowledge, the first experimentally feasible realization of
the Smolin state with continuous variables (for qubit realization see
\cite{Amselem2009_Smolin}).

\ack We are particularly thankful to Carles Rod\'o for his
valuable comments and we thank Remigiusz Augusiak for discussion.
The authors acknowledge support from the Spanish MICINN Grant No.
FIS2008-01236, Generalitat de Catalunya (SGR2009:00343) and
Consolider Ingenio 2010 (CDS2006-00019). J.S. is supported by the
Spanish Ministry of Education through the program FPU. S.P. is
supported by the Spanish Ministry of Science and Innovation
through the program Juan de la Cierva.

\section*{References}
\bibliographystyle{apsrev}
\bibliography{methods_iop}

\end{document}